\documentclass{pasj00}
\usepackage{lscape}

\begin{document}
\SetRunningHead{Kuno et al.}{Nobeyama CO Atlas of Nearby Spiral Galaxies}
\Received{2006/06/12}
\Accepted{2006/11/13}

\title{Nobeyama CO Atlas of Nearby Spiral Galaxies: Distribution of Molecular Gas in Barred and Non-barred Spiral Galaxies}

\author{Nario \textsc{Kuno},\altaffilmark{1,2} 
Naoko \textsc{Sato},\altaffilmark{3} 
Hiroyuki \textsc{Nakanishi},\altaffilmark{1} \\
Akihiko \textsc{Hirota},\altaffilmark{4} 
Tomoka \textsc{Tosaki},\altaffilmark{1} 
Yasuhiro \textsc{Shioya},\altaffilmark{5} \\
Kazuo \textsc{Sorai},\altaffilmark{6} 
Naomasa \textsc{Nakai},\altaffilmark{7} 
Kota \textsc{Nishiyama},\altaffilmark{8}
and Baltsar \textsc{Vila-Vilar\'o},\altaffilmark{9}
}
\altaffiltext{1}{Nobeyama Radio Observatory\thanks{Nobeyama Radio Observatory 
(NRO) is a division of the National Astronomical Observatory of Japan (NAOJ) under the National Institutes of Natural Sciences (NINS).}, Minamimaki-mura, Minamisaku-gun, Nagano 384-1305, Japan}
\email{kuno@nro.nao.ac.jp}
\altaffiltext{2}{The Graduate University for Advanced Studies (SOKENDAI), \\
2-21-1 Osawa, Mitaka, Tokyo 181-0015, Japan}
\altaffiltext{3}{Student Center for Independent Research in the Science, \\
Wakayama University, Wakayama 640-8510, Japan}
\altaffiltext{4}{Department of Astronomy, School of Science, The University of Tokyo, \\
 Bunkyo-ku, Tokyo 113-0033, Japan}
\altaffiltext{5}{Department of Physics, Faculty of Science, Ehime University, Matsuyama 790-8577, Japan}
\altaffiltext{6}{Division of Physics, Graduate School of Science, Hokkaido University, Sapporo 060-0810, Japan}
\altaffiltext{7}{Institute of Physics, University of Tsukuba, Ten-nodai, 1-1-1 Tsukuba, Ibaraki 305-8577, Japan}
\altaffiltext{8}{Bisei Spaceguard Center, Bisei-cho, Oda-gun, Okayama 714-1415, Japan}
\altaffiltext{9}{National Astronomical Observatory of Japan, 2-21-1 Osawa, Mitaka, Tokyo 181-0015, Japan}


\maketitle

\begin{abstract}

The data from a CO(1 -- 0) mapping survey of 40 nearby spiral galaxies performed with the Nobeyama 45-m telescope are presented. 
The criteria of the sample selection were (1) RC3 morphological type in the range Sa to Scd, (2) distance less than 25 Mpc, (3) inclination angle less than 79$^{\circ}$ (RC3), (4) flux at 100 $\mu$m higher than $\sim$ 10 Jy, (5) spiral structure is not destroyed by interaction. 
The maps of CO cover most of the optical disk of the galaxies. 
We investigated the influence of bar on the distribution of molecular gas in spiral galaxies using these data. 
We confirmed that the degree of central concentration is higher in barred spirals than in non-barred spirals as shown by the previous works. 
Furthermore, we present an observational evidence that bars are efficient in driving molecular gas that lies within the bar length toward the center, while the role in bringing gas in from the outer parts of the disks is small.
The transported gas accounts for about half of molecular gas within the central region in barred spiral galaxies. 
We found a correlation between the degree of central concentration and bar strength. 
Galaxies with stronger bars tend to have higher central concentration. 
The correlation implies that stronger bars accumulate molecular gas toward the center more efficiently. 
These results are consistent with long-lived bars.

\end{abstract}

\section{Introduction}

Since molecular gas is a key material for star formation that is a fundamental process of galaxy evolution, observations of the molecular gas are essential to understand galaxy evolution. 
CO observations are the easiest way to observe the molecular gas. 
It is difficult to observe the large-scale distribution of molecular gas in our Galaxy, because we are within the structure. 
On the other hand, we can observe the distribution of the molecular gas directly for external galaxies. 
Therefore, some CO surveys of galaxies have been done so far (e.g., Young et al. 1995; Braine et al. 1993; Nishiyama, Nakai 2001; Elfhag et al. 1996). 
These data provide important information of galaxies about global properties, such as radial distribution of molecular gas in the galactic disks. 
On the other hand, surveys carried out with interferometers revealed detailed structures of molecular gas, especially in the central region of galaxies (e.g., Sakamoto et al. 1999a; Sofue et al. 2003a; Regan et al. 2001; Helfer et al. 2003).

Although single dish telescopes are better at measuring total flux, mapping CO in whole disks of external galaxies by large single dishes has been limited so far. 
This is because a lot of time is needed to map external galaxies with a single-beam receiver with a high angular resolution. 
The Nobeyama 45-m telescope is one of the most suitable telescopes for such observations, since the 25-Beam Array Receiver System (BEARS: Sunada et al. 2000) mounted on the telescope has a high performance for mapping observations. 
Therefore, we performed a CO-mapping survey of nearby spiral galaxies with the 45-m telescope in addition to previously observed data with other receivers of the telescope. 
Although the angular resolution is poor as compared with interferometers, the single dish has an advantage of sampling total flux and our maps cover wider area than those by interferometers. 
The data is useful to investigate the relation between the distribution of molecular gas and galactic structures such as spiral arms and bars. 

We present in this paper the data of our CO (1 -- 0) mapping survey with the NRO 45-m telescope and investigate the influence of bar on the distribution of molecular gas in spiral galaxies. 
Many theoretical studies and numerical simulations have shown that a bar can play an important role for gas fueling toward the central region (Combes, Gerin 1985; Pfenniger, Norman 1990; Athanassoula 1992; Wada, Habe 1995). 
Such fueling mechanism is needed for starbursts or AGN activities in the central region. 
Recently, it has been shown observationally that the degree of central concentration of molecular gas is higher in barred spirals than non-barred spirals (Sakamoto et al. 1999b; Sheth et al. 2005). 
Furthermore, bars may play an important role on the secular evolution of galaxies by changing the distribution of molecular gas (Kormendy, Kennicutt 2004). 
Therefore, it is important to make clear how bars work on the molecular gas. We investigate the relation between distribution of molecular gas and the existence of a bar, and the correlation between the central concentration and the bar strength.

The sample criteria are described in section 2. 
The details of the observations and data reduction are described in section 3. 
The data are presented in section 4 and the influence of bars on the distribution of molecular gas is discussed in section 5. 

\section{Sample of Galaxies}

We selected our sample of galaxies according to the following criteria. 
(1) Morphological type in the RC3 (de Vaucouleurs et al. 1991) in the range from Sa to Scd. 
(2) Distance less than 25 Mpc. 
(3) The isophotal axial ratio in the RC3 ($log R$) less than 0.7. 
This value corresponds to an inclination angle of less than 79 $^{\circ}$ (most galaxies but four are $< 70^{\circ}$). 
(4) Flux density at 100 $\mu$m in the IRAS Point Source Catalogue higher than $\sim$ 10 Jy. 
(5) Spiral structure is not destroyed by interaction with other galaxies. 

We observed 40 galaxies including galaxies whose data have been already published (M51: Nakai et al. 1994; NGC 5055: Kuno et al. 1997; NGC 3504: Kuno et al. 2000; NGC 253: Sorai et al. 2000a; NGC 4212, NGC 4402, NGC 4419, NGC 4548, NGC 4689: Nakanishi 2005). 
The sample galaxies are listed in table 1. 
The distribution of the Hubble type (RC3), arm class (AC) (Elmegreen, Elmegreen 1987) and bar/non-bar classification (RC3) of the sample galaxies are listed in table 2. 

\section{Observations}

Observations of $^{12}$CO(J=1 -- 0) emission were made during 1995 to May 2005 using the 45-m telescope of the Nobeyama Radio Observatory. 
The SIS focal plane array, BEARS, which consists of 25 beams (Sunada et al. 2000), was used as a receiver front-end since December 2001. 
The beam separation of BEARS is $\sim 41$ arcsec. Digital spectrometers were used as receiver backends (Sorai et al. 2000b). 
The total bandwidth and frequency resolution of the spectrometers were 512 MHz and 605 kHz, respectively, which correspond to 1331 km s$^{-1}$ and 1.57 km s$^{-1}$ at 115 GHz. 
Since BEARS was operated in the DSB (double side band) mode, we obtained scaling factors for intensity calibration by observing a standard source with each of the 25 beams and with a single beam receiver equipped with an SSB (Single Side Band) filter, and then corrected the DSB intensity into the SSB intensity. 
We observed each point with more than 4 different receiver channels to reduce non-uniformity of the noise level due to the variation of the system noise temperature on receiver channels. 
We mapped 28 galaxies with BEARS. 
Before the operation of BEARS started, 11 galaxies were mapped with the 4-beam SIS receiver, S115Q. 
The used receiver for each galaxy is listed in table 1. 
The beam separation of S115Q was 34 arcsec. 
Since S115Q was also operated in the DSB mode, we obtained scaling factors for the intensity calibration in the same way as for BEARS. 
Only M51 was observed with single beam receivers by Nakai et al. (1994). 
For S115Q and single beam receivers, Acousto-Optical Spectrometers (AOS) were used as backends. 
The total bandwidth and frequency resolution of AOS were 250 MHz and 250 kHz, which correspond to 650 km s$^{-1}$ and 0.65 km s$^{-1}$, respectively. 

The antenna temperature, $T_{\rm A}^{*}$, was obtained by the chopper-wheel method correcting for atmospheric and antenna ohmic losses. 
The system noise temperatures during the observations were 500 K- 1000 K (SSB) in $T_{\rm A}^{*}$. 
We converted $T_{A}^{*}$ into the main beam temperature, $T_{\rm mb}$, using the main beam efficiency of $\eta_{\rm mb} = 0.4$ ($T_{\rm mb} \equiv T_{\rm A}^{*} / \eta_{\rm mb}$). 
The beam size (Full Width at Half Maximum) was about 15 arcsec at 115 GHz. 
The telescope pointing was checked every 30 min - 1 hour by observing nearby SiO maser stars or bright QSOs. 
The pointing error was less than 7 arcsec. 
We used position-switching with an integration time of 20 sec per scan and repeated many scans per position. 
The typical rms noise level is 40 -- 100 mK with the velocity resolution of 5 km s$^{-1}$. 
We adopted the position angle of the mapping grids so that the grids were parallel or perpendicular to the major axis of the galaxies taken from the literature, except for Maffei 2 whose grids were parallel and perpendicular to the bar. 
The grid spacing of most samples were 10 -- 11 arcsec except for NGC 253 and M51. 
The observational parameters are summarized in table 1.

\section{Results and Basic Parameters of Galaxies}

We checked the consistency of the intensity calibration with other observations. 
Figure 1 is the comparison between the total fluxes derived from our maps and those estimated in Young et al. (1995). 
It shows that both data are consistent.

The $J + H + K_{s}$ image from 2MASS Large Galaxy Atlas (Jarrett et al. 2003) except for NGC 4212 ($R$ band: Koopmann et al. 2001), the map of the CO integrated intensity, $I_{\rm CO} \equiv \int T_{\rm mb} dv$, the velocity field, and the position -- velocity (P -- V) diagram along the major axis of the sample galaxies are shown in figures 2 -- 41. 
Short comments on individual galaxies are also given in the figure captions.

We determined the basic parameters of the sample galaxies using the CO data and listed in table 3. 
We determined the position angle, the inclination angle and the systemic velocity kinematically using GAL in AIPS. 
The results are consistent with previous measurements except for NGC 3184. 
Inclination angles would only be derived for only a limited number of galaxies, because it is difficult to determine it precisely for most cases. 
For galaxies for which it is difficult to derive these parameters from our data, we list the values in the literature. 
We derived the mass of the molecular gas adopting the value of $1 \times 10^{20}$ cm$^{-2}$ [K km s$^{-1}$]$^{-1}$ from Nakai and Kuno (1995) as the conversion factor from the integrated intensity of the CO emission $I_{\rm CO}$ to the column density of H$_{2}$ $N$(H$_{2}$).

\section{Distribution of Molecular Gas in Barred Spiral Galaxies}

\subsection{Morphological Properties}

It has been pointed out that the radial distributions of molecular gas in barred spirals show some characteristic features (Nakai 1992; Nishiyama et al. 2001; Jogee et al. 2005). 
Most of the barred spirals have a strong intensity peak at the center, as seen in our CO maps. 
From the strong peak at the center, the slope of the radial distribution becomes shallower toward the radius of the bar-ends than the outer region. 
For some extreme cases, there is a secondary peak near the radius of the bar length. 
The same trends are seen in our maps. 
Our data supports the facts that the molecular gas distribution in barred spiral galaxies has distinctive features compared with non-barred spirals. 
Most of the barred spirals in our sample have a strong peak at the center and the molecular gas distributes along the bars (e.g., IC 342, Maffei 2). 
The ridges of the molecular gas along the bars offset toward the leading side as shown by many numerical simulations (e.g., Athanassoula 1992) and previous observations (e.g., Sheth et al. 2002). 
In some galaxies (e.g., Maffei 2, NGC 2903, NGC 3627, NGC 6951), the bar-ends are the places where molecular gas tends to concentrate. 
On the other hand, the CO-intensity peak at the center in non-barred spirals except for NGC 5055 is not so prominent. 
Some galaxies show kiloparsec-scale central gas depletion (NGC 4212, NGC 4414, NGC 4689). 
It is interesting that these galaxies are all flocculent type. 
Another flocculent type NGC 3521 also has central depletion, although the galaxy is classified as SAB. 
Besides our sample, NGC 2841 and NGC 7331 have been known to have central depletion (Young, Scoville 1982). 
Both galaxies are also flocculent type. 
These results may give a clue to understanding the cause of the kiloparsec-scale central depletion. 
We will present the detail radial distributions of molecular gas in our sample galaxies and discuss their relation with rotation curves in a forthcoming paper. 

\subsection{Central Concentration of Molecular Gas}
\subsubsection{Indicators of Central Concentration of Molecular Gas}

Sakamoto et al. (1999b) and Sheth et al. (2005) showed that the degree of central concentration of molecular gas is higher in barred spirals than in non-barred spirals, using interferometric survey data. 
Here, we also investigate the degree of central concentration of molecular gas in our sample galaxies. 
The difference from the previous works is that we normalize the areas where we derive molecular gas mass with the size of galaxies. 
This is important because if we adopt a fixed absolute radius as the central region as Sakamoto et al. (1999b) and Sheth et al. (2005), there is a possibility that the dependence on the galaxy size appears. 
Furthermore, the normalized scale makes it possible to make more detail comparisons. 
In order to show the role of bars clearer, we compare the accumulations of molecular gas within the bar and from the outside of the bar region using two indicators. 
The definition of the indicators, $f_{\rm in}$ and $f_{\rm out}$, are as follows.
\begin{equation}
    f_{\rm in} = \frac{M_{\rm H_{2}}(R_{\rm K20}/8)}{M_{\rm H_{2}}(R_{\rm K20}/2)}
\end{equation}
\begin{equation}
    f_{\rm out} = \frac{M_{\rm H_{2}}(R_{\rm K20}/2)}{M_{\rm H_{2}}(R_{\rm K20})},
\end{equation}
where $M_{\rm H_{2}}(R)$ is molecular gas mass within a radius $R$ and $R_{\rm K20}$ is the radius at the 20 mag arcsec$^{-2}$ in the $K_{S}$-band. 
We used the $R_{\rm K20}$ from the 2MASS Large Galaxy Atlas (Jarrett et al. 2003). 

We measured the semi-major length of bar, $R_{\rm bar}$, using the 2MASS $K_{s}$ images for 18 galaxies of our sample whose bars can be clearly seen in the image (table 3). 
We define the radius where the amplitude and phase of the $m = 2$ Fourier component change significantly as the bar length. 
The average and the maximum of the ratio of $R_{\rm bar}$ to $R_{K20}$ are 0.37 and 0.59, respectively (figure 42). 
We selected the radius of $R_{\rm K20}/2$ so that $f_{\rm in}$ highlights the degree of the concentration toward the central region ($R < R_{\rm K20}/8$) within radial distances of the order of a bar length. 
The size of the central region is preferred to be as compact as possible, but should be larger than the beam size of 15 arcsec. 
Otherwise, smaller galaxies may show lower central concentration due to beam smearing. 
For these requirements, we defined $R_{\rm K20}/8$ as the central region. 
On the other hand, $f_{\rm out}$ represents the fraction of molecular gas mass within the radius of the order of a bar length ($R < R_{\rm K20}/2$). 
We derive $M_{\rm H_{2}}(R)$ from the re-gridded CO maps assuming that the conversion factor from $I_{\rm CO}$ to H$_{2}$ column density is constant within each galaxy. 
Therefore, $f_{\rm in}$ and $f_{\rm out}$ do not depend on the conversion factor. 
$R_{\rm K20}/2$ is in the range of 33 arcsec to 315 arcsec, and the average is 77 arcsec. 
$R_{\rm K20}/2$ in the linear scale is 1.8 kpc to 6.8 kpc, and the average is 4.1 kpc. 
On the other hand, $R_{\rm K20}$/8 of our sample is 8.2 arcsec to 78.8 arcsec, and the average is 19.3 arcsec. 
The minimum is still larger than half of our beam size (7.5 arcsec). 
$R_{\rm K20}$/8 in the linear scale is 0.4 kpc to 1.7 kpc, and the average is 1.0 kpc. 

\subsubsection{Central Concentration within Bar Region}

Figure 43 shows the distribution of $f_{\rm in}$ in barred (SAB+SB) and non-barred spirals (SA). 
The averages of $f_{\rm in}$ for non-barred and barred spirals are $0.13 \pm 0.03$ and $0.20 \pm 0.08$, respectively, where we note that the uniform distribution gives $f_{\rm in} = (2/8)^{2} = 0.06$. 
Thus, barred spirals have higher degree of the central concentration and larger dispersion than non-barred spirals. 
This result is consistent with the previous works by Sakamoto et al. (1999b) and Sheth et al. (2005). 
The probability that both samples are extracted from same parent sample (the Kolmogorov-Smirnov test) is 0.004. 
It is confirmed in figure 44 that the result is not due to the bias of the apparent size of the galaxies. 
The trend is clearly seen in all range of the apparent size. 
One may think that a bias of morphological type may cause the difference. Figure 45 shows the relation between $f_{\rm in}$ and Hubble type. 
The dependence of $f_{\rm in}$ on the Hubble type is not apparent and non-barred galaxies distribute near the bottom for all type. Sheth et al. (2005) showed that most of the highest central concentrations are seen in CO-bright galaxies. 
Since we selected sample galaxies using the flux density at 100 $\mu$m $F_{\rm 100\mu m}$ to select CO bright galaxies, we checked the dependence on $F_{\rm 100\mu m}$ (figure 46). 
As pointed out by Sheth et al. (2005), half of the galaxies with high 100 $\mu$m flux ($> 100$ Jy) shows large $f_{\rm in}$ ($> 0.25$). 
However, it is also confirmed that the trend that the central concentration is higher in barred spirals than non-barred spirals is seen regardless of $F_{\rm 100\mu m}$.

We estimate the mass of the molecular gas transported by the bar toward the central region ($R < R_{\rm K20}/8$), although it depends on the size of a galaxy and the total mass of molecular gas. 
Using the average of $f_{\rm in}$ of SA (0.13), we assume that $M_{\rm H_{2}}(R_{\rm K20}/8)$ is $0.13 \times M_{\rm H_{2}}(R_{\rm K20}/2)$ if the galaxy does not have a bar. 
Then, we regard the excess of the observed $M_{\rm H_{2}}(R_{\rm K20}/8)$ of barred galaxies from the estimation for non-barred as the transported gas by the bar. 
For galaxies whose $f_{\rm in}$ is larger than 0.16 (the average of SA plus 1$\sigma$), the derived mass is in the range of $6.9 \times 10^{6} M_{\odot}$ to $2.3 \times 10^{8} M_{\odot}$ and the average is $1.1 \times 10^{8} M_{\odot}$. 
It is difficult to estimate the amount of the gas consumed by star formation until now, since star formation rate changes depending on the amounts of molecular gas. 
If, however, we assume that star formation efficiency is $2.4 \times 10^{-9}$ yr$^{-1}$, which is the mean value among the face-on undisturbed galaxies derived in Rownd and Young (1999), and simply that the present molecular gas mass $M_{\rm H_{2}}$ is the average value during the galaxy evolution, the molecular gas consumed by star formation in $10^{10}$yr is estimated to be 24 $ \times M_{\rm H_{2}}$. 
This is an upper limit, if the inflow rate is higher than star formation rate (Sakamoto et al. 1999b). 
Therefore, the total amount of the transported molecular gas is estimated to be $2.8 \times 10^{9} M_{\odot}$. 
If we assume that a rotation velocity of 200 km s$^{-1}$ at $R = 1$ kpc, for example, the dynamical mass within the radius is $\sim 1 \times 10^{10} M_{\odot}$, and therefore, the transported mass corresponds to about 30\% of the dynamical mass. 
We have to note that the value involves a large uncertainty from the estimations of the star formation efficiency, the average molecular gas mass and the CO-H$_{2}$ conversion factor. 
On the other hand, the proportion of the transported gas to the observed $M_{\rm H_{2}}(R_{\rm K20}/8)$ is not altered by them, if we assume that a certain fraction of molecular gas is consumed by star formation regardless whether barred or non-barred. 
The proportion is 0.36 to 0.65, and the average is 0.49. 
Therefore, about half of the molecular gas in the central region ($R < R_{\rm K20}/8$) may have been transported by the bar in these galaxies.

\subsubsection{Inflow from the Outside of the Bar Region}

Our data which cover wide area in the galactic discs are useful to investigate the influence of bars on the outside of the bar region. 
Figure 47 shows the distribution of $f_{\rm out}$ in barred and non-barred spirals. 
Similar distributions of $f_{\rm out}$ are seen for both groups. 
The averages of $f_{\rm out}$ of non-barred and barred spirals are $0.60 \pm 0.12$ and $0.59 \pm 0.16$, respectively (The probability of the Kolmogorov-Smirnov test is 0.36). 
This result implies that the degree of the gas concentration toward the region within radial distances of the order of the bar is comparable in barred and non-barred spirals and that bars play a little role at larger spatial scales than bars on the disks, while bars are efficient in driving molecular gas toward the galactic center. 
This result seems to be reasonable, since transformation of angular momentum of molecular gas occurs in the bar region. 
With these results, the characteristic features of the radial distribution of barred spirals can also be explained. 
Namely, because of the gas flow, the central surface density increases and the radial distribution becomes shallow within the bar region, while the distribution out of the bar is similar to that in non-barred spirals (fig. 48). 
Furthermore, our result is consistent with the evolutionary sequence proposed by Jogee et al. (2005) and Sheth et al. (2005). 
They suggested that barred spirals evolve from type I, in which molecular gas distributes in the bar region and molecular gas is been accumulating toward the central region, to type II, in which most of molecular gas is concentrated toward the central region, and then, type III, in which all of the molecular gas in the central region has been consumed by a starburst. 
Since inflow of molecular gas from the outside of the bar is not significant as shown by our result, this scenario seems to be reasonable.

\subsection{Correlation between Central Concentration and Bar Strength}

If bars accumulate molecular gas toward the center, it is expected that properties of the bars, especially the bar strength, correlate with the degree of central concentration of the gas. 
Figure 49 shows the correlation between the bar strength and the degree of central concentration ($f_{\rm in}$), where the bar strengths $Q_{H}$ were adopted from Laurikainen and Salo (2002). 
They derived the maxima of the tangential force and the average of the radial force at each radius in a bar, and then defined the maximum of the ratio between them as the bar strength. 
19 galaxies in our sample are included in their sample. 
They are plotted in figure 49. 
It is apparent that galaxies with stronger bar have higher central concentration. 
The correlation coefficient is 0.69. 
This correlation can be naturally understood, since strong bars with strong non-axisymmetric forces accumulate molecular gas more efficiently toward the central region. 
This is consistent with the results of the numerical simulations (e.g., Regan, Teuben 2004). 
Since, however, sample number is still small, it is not conclusive. 
Note that the figure indicates that the large dispersion of $f_{\rm in}$ of barred spirals shown in figure 43 is caused by the variation of the bar strength. 

Vila-Costas and Edmunds (1992) found that the global abundance gradients of barred spirals are in general shallower than that of non-barred spirals. 
For NGC 6946, the abundance gradient is apparently shallower in the bar region than in the outer region (Roy, Belley 1993). 
Furthermore, Martin and Roy (1994) found that the gradients are flatter when the length or the ellipticity of the bar increases. 
These facts are interpreted by an idea that stronger bars can accumulate molecular gas toward central region more efficiently and the shallower abundance gradient in stronger barred spirals is caused by large-scale mixing of the interstellar gas by bars (Martin, Roy 1994; Friedli, Benz 1993; Noguchi 1998). 
By comparing these results with our data, we can examine the idea. 
If it is correct, we should be able to find the correlation between central concentration and abundance gradient. 
Figure 50 shows the comparison of $f_{\rm in}$ with the abundance gradient for objects included in Martin and Roy (1994). 
The trend that galaxies with higher central concentration have shallower gradient of the abundance can be seen in the figure, being consistent with the scenario. 
Since number of galaxies in our sample whose abundance gradient has been reported is still small, we need to measure the abundance gradient for others in our sample to confirm this trend.

It is still an open question whether bars are short-lived due to destruction by mass transport toward the center or not (e.g., Bournaud, Combes 2002; Das et al. 2003; Shen, Sellwood 2004; Bournaud et al. 2005; Athanassoula et al. 2005). 
The observational results that the central concentrations are found in barred spirals (e.g., Sakamoto et al. 1999b) support long-lived bars. Furthermore, bars are found in galaxies at $z \sim 1$ (Sheth et al. 2003) and it is shown that the bar fraction is constant with time up to $z \sim 1$ (Elmegreen et al. 2004), although more sample galaxies are needed to get conclusive evidence. 
Our finding of the correlation between the bar strength and the degree of central concentration also seems to support the long-lived bars, although it is not direct evidence. 
Such correlation must be hardly seen if bars are dissolved by the gas accumulation and short-lived, since the central concentration weakens the bar strength. 
If bars are destroyed in 1 -- 2 Gyr (Bournaud, Combes, Semelin 2005), bars must be reformed to maintain the constant fraction of barred spirals with time. 
In that case, the bar strength oscillates because of its destruction and reformation, while the central concentration grows with time (Bournaud, Combes 2002). 
As a result, the correlation between the bar strength and the central concentration dilutes. 
Therefore, these results seem to be consistent with long-lived bars rather than short-lived ones which are destroyed by gas accumulation toward the center many times in the Hubble time.

\section{Summary}

We made a CO(1 -- 0) mapping survey of 40 nearby spiral galaxies with the Nobeyama 45-m telescope to provide useful data for detailed and systematic studies of molecular gas in the galaxies. 
Using these data we have compared the distribution of molecular gas in barred and non-barred spirals and investigated the influence of the bar. 
We confirmed that the degree of the central concentration of molecular gas within the radius of the order of a bar length ($f_{\rm in}$) in barred spirals is significantly higher than that in non-barred spirals as shown by Sakamoto et al. (1999b) and Sheth et al. (2005). 
This is contrast with the degree of the concentration of the total molecular gas mass within the radial distances of the order of the bar ($f_{\rm out}$), which is similar for both barred and non-barred spirals. 
This implies that the bars appear to be efficient in driving gas that lies within their radial scales toward the center of the host galaxies, but that they play quite a smaller role at larger spatial scales on the disks. 
Thus the characteristic feature of the radial distribution of molecular gas seen in barred spirals, i.e. the strong intensity peaks at their centers, the shallow gradients within the bar regions or/and the secondary peaks at the radius of the bar-ends, can be explained by the accumulation of molecular gas within the bar regions. 
The accumulated gas by bars accounts for about half of molecular gas within the central region. 
We also found a correlation between the degree of central concentration of molecular gas, $f_{\rm in}$, and the bar strength. 
Galaxies with stronger bars tend to have higher central concentrations. 
A correlation between the degree of central concentrations of molecular gas and the abundance gradient of heavy elements was also found. 
Galaxies with higher $f_{\rm in}$ have shallower abundance gradient. 
These results indicate that stronger bar accumulate molecular gas toward the center more efficiently. 
The correlation between the degree of central concentration of molecular gas and the strength seem to be consistent with long-lived bars rather than short-lived ones which are destroyed by the gas accumulation toward the center many times in the Hubble time.

\bigskip

We thank the NRO staff for their kind support and continuous encouragements. We also thank an anonymous referee for useful comments. This research has made use of the NASA/IPAC Extragalactic Database (NED) and the NASA/ IPAC Infrared Science Archive which are operated by the Jet Propulsion Laboratory, California Institute of Technology, under contract with the National Aeronautics and Space Administration. This research has made use of the VizieR catalogue access tool, CDS, Strasbourg, France.


\begin{figure}
\begin{center}

\FigureFile(80mm,50mm){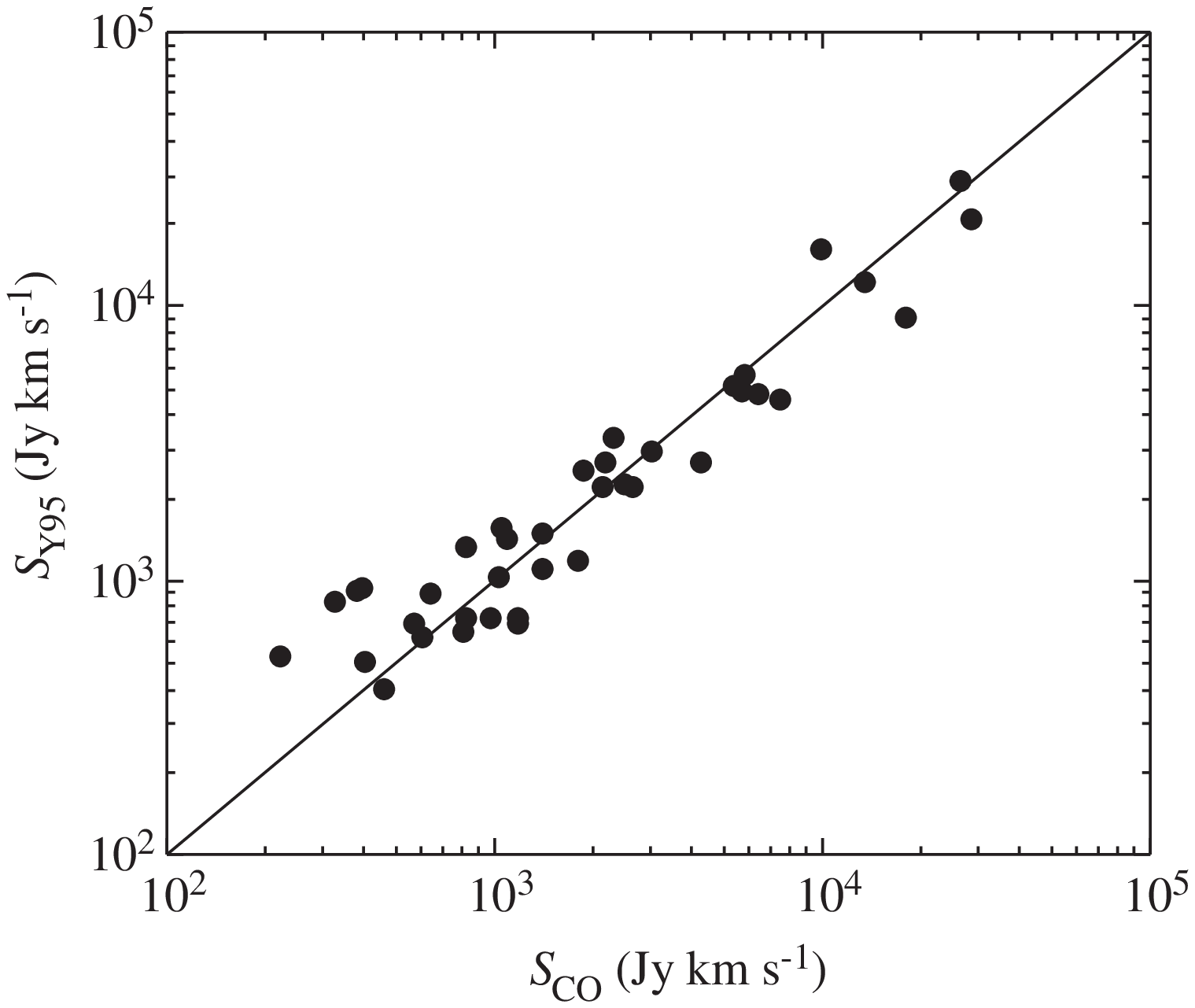}
\end{center}
\caption{Comparison of the total flux in this work with Young et al. (1995). The abscissa $S_{\rm CO}$ is the total flux derived in this work and the ordinate $S_{\rm Y95}$ is the total flux in Young et al. (1995). The solid line indicates $S_{\rm CO} = S_{\rm Y95}$.}
\label{}

\end{figure}

\begin{figure}
\begin{center}
\end{center}
\caption{ NGC 253. Top-left is the $J + H + K$ image from 2MASS Large Galaxy Atlas. Top-right is the CO integrated intensity map. Contours levels are 10 $\times$ (1, 2, 3, 4, 6, 8, 10, 15, 20, 25, 30, 40, 50) K km s$^{-1}$. Bottom-left is the velocity field measured in CO. Contours levels are from 70 to 420 km s$^{-1}$ with an interval of 30 km s$^{-1}$. Bottom-right is the position -- velocity diagram of CO along the major axis. Contours levels are 0.2 $\times$ (1, 2, 3, 4, 6, 8, 10) K. The data were taken from Sorai et al. (2000a). The central peak and the ring of molecular gas surrounding the bar are prominent.
}
\label{}
\end{figure}

\begin{figure}
\begin{center}
\end{center}
\caption{Same as figure 2 for Maffei 2. Contour levels of the CO integrated intensity map are 6 $\times$ (1, 2, 3, 4, 6, 8, 10, 15, 20, 25, 30, 35, 40) K km s$^{-1}$. Contour levels of the velocity field are from -180 to 120 km s$^{-1}$ with an interval of 30 km s$^{-1}$. Contour levels of the P -- V diagram are 0.1 $\times$ (1, 2, 3, 4, 6, 8, 10, 15) K. Mason and Wilson (2005) have shown that there are strong concentrations of molecular gas at the bar-ends along with the center. Our CO map shows strong peak at the center, offset ridges along the leading side of the bar, condensations at the bar ends, and spiral arms. These are structures often seen in barred spiral galaxies. There are clear gaps between offset ridges and spiral arms. The gaps are often seen in the galaxies that have straight dust lanes along the bar. NGC 2903, NGC 6951, and IC 342 are other examples. Large non-circular motion is seen in the bar region. Although the optical and the near-IR images show disturbed feature, the CO map and velocity field show symmetric structure. Since Maffei 2 locates behind the Milky Way, large extinction even in the near-IR band seems to be the cause of the asymmetric feature.
}
\label{}

\end{figure}

\begin{figure}
\begin{center}
\end{center}
\caption{Same as figure 2 for NGC 1068. Contour levels of the CO integrated intensity map are 5 $\times$ (1, 2, 3, 4, 6, 8, 10, 15, 20, 25) K km s$^{-1}$. Contour levels of the velocity field are from 1020 to 1260 km s$^{-1}$ with an interval of 20 km s$^{-1}$. Contour levels of the P -- V diagram are 0.15 $\times$ (1,2,3,4,6) K. Molecular gas concentrates within $R < 1$ arcmin. The spiral structure seen in the CO map by BIMA (Helfer et al. 2003) seems to be buried in the extended component and corresponds to the ridge in our map. 
}
\label{}

\end{figure}

\begin{figure}
\begin{center}

\end{center}
\caption{Same as figure 2 for IC 342. Contour levels of the CO integrated intensity map are 6 $\times$ (1, 2, 3, 4, 6, 8, 10, 15, 20, 25, 30) K km s$^{-1}$. Contour levels of the velocity field are from -50 to 110 km s$^{-1}$ with an interval of 10 km s$^{-1}$. Contour levels of the P -- V diagram are 0.2 $\times$ (1, 2, 3, 4, 6, 8, 10, 15) K. Since IC 342 is located behind the Milky Way, the optical morphology is not so clear. The 2 MASS image (Jarrett et al. 2003) shows that the galaxy has a large bar whose length of the semi-major axis is $\sim$ 120 arcsec and position angle is about 155 deg. Our CO map shows the offset ridges along the leading side of the bar and concentrations at the bar-ends, which are not clear in the previous maps with lower angular resolution (Sage, Solomon 1991; Crosthwaite et al. 2001). The southern arm that starts from the southern bar-end is much stronger than the northern arm that starts from the northern bar-end in the CO map. This is the same trend seen in the 20-cm radio continuum (Condon 1987). Although IC 342 is classified as a weak bar SAB, the velocity field shows large non-circular motion in the bar region. Sato (2006) derived the pattern speed of the spiral structure by comparing our CO map and the 2MASS $K_{s}$ image which traces the spiral potential.
}

\label{}
\end{figure}

\begin{figure}
\begin{center}

\end{center}

\caption{Same as figure 2 for UGC 2855. Contour levels of the CO integrated intensity map are 4 $\times$ (1, 2, 3, 4, 6, 8, 10, 15) K km s$^{-1}$. Contour levels of the velocity field are from 1030 to 1390 km s$^{-1}$ with an interval of 30 km s$^{-1}$. Contour levels of the P -- V diagram are 0.15 $\times$ (1, 2, 3) K. The P-V diagram along the major axis shows the rigid-like rotation curve in the bar region. It can be seen more clearly in the high resolution interferometric map (H\"uttemeister et al. 1999). This is because the major axes of the bar and of the galaxy itself are almost parallel. For this configuration, the rigid rotation velocity which corresponds to the pattern speed of the bar is expected to be observed along the bar (Kuno et al. 2000).
}
\label{}
\end{figure}

\begin{figure}

\begin{center}
\end{center}
\caption{Same as figure 2 for NGC 2903. Contour levels of the CO integrated intensity map are 5 $\times$ (1, 2, 3, 4, 6, 8, 10, 15) K km s$^{-1}$. Contour levels of the velocity field are from 370 to 730 km s$^{-1}$ with an interval of 30 km s$^{-1}$. Contour levels of the P -- V diagram are 0.15 $\times$ (1, 2, 3, 4) K. CO emission distributes along the bar and spiral arms. There is a strong peak at the center and secondary peaks are seen at the bar ends. The molecular arms continue toward the opposite bar-end and make a ring-like structure. The velocity width in the bar region is very wide ($\sim$ 100 km s$^{-1}$). This is attributed to large velocity change due to shock at the ridge. Since the position angles of the line of nodes and the bar are close, the velocity change can be observed directly.
}
\label{}
\end{figure}

\begin{figure}

\begin{center}

\end{center}
\caption{Same as figure 2 for NGC 3184. Contour levels of the CO integrated intensity map are 2 $\times$ (1, 2, 3, 4, 5, 6) K km s$^{-1}$. Contour levels of the velocity field are from 550 to 630 km s$^{-1}$ with an interval of 10 km s$^{-1}$. Contour levels of the P -- V diagram are 0.12 $\times$ (1, 2, 3) K. Molecular gas traces the two-arm spiral structure seen in the NIR image. The molecular arms are clumpy. As seen in the velocity field, it is apparent that the position angle of the major axis taken from RC3 (45 deg) is incorrect.
}
\label{}
\end{figure}

\begin{figure}
\begin{center}
\end{center}
\caption{Same as figure 2 for NGC 3351. Contour levels of the CO integrated intensity map are 3 $\times$ (1, 2, 3, 4, 6, 8, 10, 15) K km s$^{-1}$. Contour levels of the velocity field are from 660 to 920 km s$^{-1}$ with an interval of 20 km s$^{-1}$. Contour levels of the P -- V diagram are 0.1 $\times$ (1, 2, 3, 4) K. Strong concentration of molecular gas is seen in the center. In addition to the central peak, some clumps of molecular gas distribute along a ring surrounding the bar. The molecular gas is depleted in the bar region.
}
\label{}
\end{figure}

\begin{figure}
\begin{center}
\end{center}
\caption{Same as figure 2 for NGC 3504. Contour levels of the CO integrated intensity map are 3 $\times$ (1, 2, 3, 4, 6, 8, 10, 15) K km s$^{-1}$. Contour levels of the velocity field are from 1460 to 1640 km s$^{-1}$ with an interval of 20 km s$^{-1}$. Contour levels of the P -- V diagram are 0.05 $\times$ (1, 2, 3, 4, 6) K. The data were taken from Kuno et al. (2000). There is a strong peak at the center. Offset ridges of the molecular gas along the bar which continue to the bar end smoothly can be seen. On the other hand, the spiral arms are weak in CO as compared with the bar region. Some galaxies that have curved dust lanes along the bar have this trend. For example, NGC 4321 and NGC 5248 seem to belong to this type.
}
\label{}
\end{figure}

\begin{figure}
\begin{center}
\end{center}
\caption{Same as figure 2 for NGC 3521. Contour levels of the CO integrated intensity map are 8 $\times$ (1, 2, 3, 4, 6) K km s$^{-1}$. Contour levels of the velocity field are from 600 to 990 km s$^{-1}$ with an interval of 30 km s$^{-1}$. Contour levels of the P -- V diagram are 0.15 $\times$ (1, 2, 3, 4) K. NGC 3521 is one of the galaxies which have a kiloparsec-scale central depletion of molecular gas (Nishiyama et al. 2001; Helfer et al. 2003). Our map also clearly shows the depletion at the center.
}
\label{}
\end{figure}

\clearpage

\begin{figure}
\begin{center}

\end{center}
\caption{Same as figure 2 for NGC 3627. Contour levels of the CO integrated intensity map are 5 $\times$ (1, 2, 3, 4, 6, 8, 10, 15, 20) K km s$^{-1}$. Contour levels of the velocity field are from 550 to 910 km s$^{-1}$ with an interval of 30 km s$^{-1}$. Contour levels of the P -- V diagram are 0.15 $\times$ (1, 2, 3, 4) K. Molecular gas concentrates at the center and bar ends. Spiral arms are also traced by CO. The asymmetric feature of the arms caused by the interaction with NGC 3628 (Zhang et al. 1993) is seen in the distribution of molecular gas. The velocity width in the bar region is large as seen in NGC 2903. 
}
\label{}

\end{figure}

\begin{figure}
\begin{center}

\end{center}
\caption{Same as figure 2 for NGC 3631. Contour levels of the CO integrated intensity map are 3 $\times$ (1, 2, 3, 4, 6, 8) K km s$^{-1}$. Contour levels of the velocity field are from 1120 to 1200 km s$^{-1}$ with an interval of 20 km s$^{-1}$. Contour levels of the P -- V diagram are 0.1 $\times$ (1, 2, 3, 4) K. The central peak is prominent. Weak spiral arms are seen in the CO map.
}
\label{}

\end{figure}

\begin{figure}
\begin{center}
\end{center}
\caption{Same as figure 2 for NGC 4051. Contour levels of the CO integrated intensity map are 5 $\times$ (1, 2, 3, 4, 6) K km s$^{-1}$. Contour levels of the velocity field are from 620 to 780 km s$^{-1}$ with an interval of 20 km s$^{-1}$. Contour levels of the P -- V diagram are 0.1 and 0.2 K. Molecular gas concentrates toward the center and the bar ends. 
}\label{}
\end{figure}

\begin{figure}
\begin{center}
\end{center}
\caption{Same as figure 2 for NGC 4102. Contour levels of the CO integrated intensity map are 5 $\times$ (1, 2, 3, 4, 6, 8, 10, 15, 20) K km s$^{-1}$. Contour levels of the velocity field are from 700 to 970 km s$^{-1}$ with an interval of 30 km s$^{-1}$. Contour levels of the P -- V diagram are 0.1 $\times$ (1, 2, 3, 4) K. Molecular gas concentrates toward the center. The rotation curve rises steeply at the center.
}\label{}
\end{figure}

\begin{figure}

\begin{center}
\end{center}
\caption{Same as figure 2 for NGC 4192. Contour levels of the CO integrated intensity map are 4 $\times$ (1, 2, 3) K km s$^{-1}$. Contour levels of the velocity field are from -360 to 120 km s$^{-1}$ with an interval of 40 km s$^{-1}$. Contour levels of the P -- V diagram are 0.1 $\times$ (1, 2, 3) K. Since the inclination angle is large, the distribution of the molecular gas is along the major axis.
}\label{}
\end{figure}

\begin{figure}
\begin{center}
\end{center}
\caption{Same as figure 2 for NGC 4212. Top-left is $R$-band image. Contour levels of the CO integrated intensity map are 3 $\times$ (1, 2, 3, 4) K km s$^{-1}$. Contour levels of the velocity field are from -180 to 20 km s$^{-1}$ with an interval of 20 km s$^{-1}$. Contour levels of the P -- V diagram are 0.07 $\times$ (1, 2, 3) K. Kiloparsec-scale central depletion of molecular gas is seen. The rotation curve shows rigid-like feature within the ring-like structure. 
}
\label{}
\end{figure}

\begin{figure}
\begin{center}
\end{center}
\caption{Same as figure 2 for NGC 4254. Contour levels of the CO integrated intensity map are 5 $\times$ (1, 2, 3, 4, 6, 8) K km s$^{-1}$. Contour levels of the velocity field are from 2300 to 2480 km s$^{-1}$ with an interval of 20 km s$^{-1}$. Contour levels of the P-V diagram are 0.1 $\times$ (1, 2, 3, 4, 6, 8) K. A central spiral structure of molecular gas has been revealed with interferometers (Sakamoto et al. 1999a; Sofue et al. 2003a). Our map shows the extension of the molecular arms beyond the field of view of the interferometers.
}

\label{}
\end{figure}

\begin{figure}
\begin{center}
\end{center}
\caption{Same as figure 2 for NGC 4303. Contour levels of the CO integrated intensity map are 10 $\times$ (1, 2, 3, 4, 6) K km s$^{-1}$. Contour levels of the velocity field are from 1510 to 1630 km s$^{-1}$ with an interval of 20 km s$^{-1}$. Contour levels of the P -- V diagram are 0.15 $\times$ (1, 2, 3, 4) K. The molecular gas distributes along the bar and spiral arms from the center. The offset ridge has been clearly resolved with interferometers (Schinnerer et al. 2002; Koda, Sofue 2006).
}\label{}
\end{figure}

\begin{figure}
\begin{center}
\end{center}
\caption{Same as figure 2 for NGC 4321. Contour levels of the CO integrated intensity map are 5 $\times$ (1, 2, 3, 4, 6, 8, 10) K km s$^{-1}$. Contour levels of the velocity field are from 1470 to 1670 km s$^{-1}$ with an interval of 20 km s$^{-1}$. Contour levels of the P -- V diagram are 0.07 $\times$ (1, 2, 3, 4, 6, 8, 10) K. The molecular gas is abundant in the center and the bar. Our map is consistent with the map obtained with IRAM 30-m telescope by Sempere and Garc\'ia-Burillo (1997).
}\label{}
\end{figure}

\begin{figure}
\begin{center}
\end{center}
\caption{Same as figure 2 for NGC 4402. Contour levels of the CO integrated intensity map are 5 $\times$ (1, 2, 3, 4, 6) K km s$^{-1}$. Contour levels of the velocity field are from 150 to 330 km s$^{-1}$ with an interval of 30 km s$^{-1}$. Contour levels of the P -- V diagram are 0.1 $\times$ (1, 2, 3, 4) K. The extents of molecular and atomic gas are almost the same (Nakanishi 2005). Since the inclination angle is large, it is difficult to know the distribution of molecular gas in the galactic disk.
}\label{}
\end{figure}

\clearpage

\begin{figure}
\begin{center}
\end{center}
\caption{Same as figure 2 for NGC 4414. Contour levels of the CO integrated intensity map are 4 $\times$ (1, 2, 3, 4, 6, 8, 10, 15) K km s$^{-1}$. Contour levels of the velocity field are from 520 to 880 km s$^{-1}$ with an interval of 30 km s$^{-1}$. Contour levels of the P -- V diagram are 0.15 $\times$ (1, 2, 3, 4, 6) K. It has been reported that NGC 4414 has the central depletion of molecular gas (Sakamoto 1996; Helfer et al. 2003). Although the central hole is not apparent in our CO map, the peak is located on the ring of molecular gas shown in the previous observations with interferometers.
}\label{}
\end{figure}

\begin{figure}
\begin{center}
\end{center}
\caption{Same as figure 2 for NGC 4419. Contour levels of the CO integrated intensity map are 4 $\times$ (1, 2, 3, 4, 6, 8) K km s$^{-1}$. Contour levels of the velocity field are from -320 to -40 km s$^{-1}$ with an interval of 40 km s$^{-1}$. Contour levels of the P -- V diagram are 0.1 $\times$ (1, 2, 3) K. The extent of molecular gas is compact at the center. Atomic gas is also seen only near the center (Nakanishi 2005).
}\label{}
\end{figure}

\begin{figure}
\begin{center}
\end{center}
\caption{Same as figure 2 for NGC 4501. Contour levels of the CO integrated intensity map are 4 $\times$ (1, 2, 3, 4, 6, 8) K km s$^{-1}$. Contour levels of the velocity field are from 2030 to 2470 km s$^{-1}$ with an interval of 40 km s$^{-1}$. Contour levels of the P -- V diagram are 0.07 $\times$ (1, 2, 3, 4) K. Onodera et al. (2004) shows that spiral arms continue to the center. Our CO velocity field shows large-scale distortion. The HI velocity field also shows the same distortion (Cayatte et al. 1990). Since HI shows a highly asymmetric distribution caused by compression by the intergalactic medium, the distortion may also be due to this compression.
}\label{}
\end{figure}

\begin{figure}
\begin{center}
\end{center}
\caption{Same as figure 2 for NGC 4535. Contour levels of the CO integrated intensity map are 3 $\times$ (1, 2, 3, 4, 6, 8, 10) K km s$^{-1}$. Contour levels of the velocity field are from 1860 to 2070 km s$^{-1}$ with an interval of 30 km s$^{-1}$. Contour levels of the P -- V diagram are 0.08 $\times$ (1, 2, 3) K. Both spiral arms are traced by the CO map. The molecular arms consist of some clumps and are not very smooth.
}\label{}
\end{figure}

\begin{figure}
\begin{center}
\end{center}
\caption{Same as figure 2 for NGC 4536. Contour levels of the CO integrated intensity map are 3 $\times$ (1, 2, 3, 4, 6, 8, 10, 15) K km s$^{-1}$. Contour levels of the velocity field are from 1660 to 1930 km s$^{-1}$ with an interval of 30 km s$^{-1}$. Contour levels of the P -- V diagram are 0.1 $\times$ (1, 2, 3) K. Molecular gas concentrates toward the center. The spiral arms are not so strong in CO except for the southeastern part of the arm.
}
\label{}
\end{figure}

\begin{figure}
\begin{center}
\end{center}
\caption{Same as figure 2 for NGC 4548. Contour levels of the CO integrated intensity map are 2 $\times$ (1, 2, 3, 4) K km s$^{-1}$. Contour levels of the velocity field are from 440 to 520 km s$^{-1}$ with an interval of 30 km s$^{-1}$. Contour levels of the P -- V diagram are 0.08 K. Weak arms along the near-IR arms are seen in the CO map. Molecular gas is depleted in the bar region.
}
\label{}
\end{figure}

\begin{figure}
\begin{center}
\end{center}
\caption{Same as figure 2 for NGC 4569. Contour levels of the CO integrated intensity map are 4 $\times$ (1, 2, 3, 4, 6, 8, 10, 15) K km s$^{-1}$. Contour levels of the velocity field are from -360 to -40 km s$^{-1}$ with an interval of 40 km s$^{-1}$. Contour levels of the P -- V diagram are 0.1 $\times$ (1, 2, 3, 4) K. Concentrations of molecular gas are seen at the center and the bar ends. The P-V diagram is slightly asymmetric. The velocity width at the northern side is larger than the southern side. This is consistent with previous interferometric observations (Nakanishi et al. 2005). The central peak is resolved into a circumnuclear elliptical ring by the interferometer.
}
\label{}
\end{figure}

\begin{figure}

\begin{center}
\end{center}
\caption{Same as figure 2 for NGC 4579. Contour levels of the CO integrated intensity map are 3 $\times$ (1, 2, 3, 4) K km s$^{-1}$. Contour levels of the velocity field are from 1370 to 1670 km s$^{-1}$ with an interval of 30 km s$^{-1}$. Contour levels of the P -- V diagram are 0.06 and 0.12 K. Molecular gas distributes along the spiral arms. The molecular arms have not been detected with interferometers because of the limited field of view (Sofue et al. 2003a; Helfer et al. 2003). The central peak is not prominent.
}
\label{}

\end{figure}

\begin{figure}
\begin{center}
\end{center}
\caption{Same as figure 2 for NGC 4654. Contour levels of the CO integrated intensity map are 4 $\times$ (1, 2, 3, 4, 6) K km s$^{-1}$. Contour levels of the velocity field are from 930 to 1170 km s$^{-1}$ with an interval of 30 km s$^{-1}$. Contour levels of the P -- V diagram are 0.1 $\times$ (1, 2, 3) K. The northern arm is stronger than the southern arm in H$\alpha$ maps (Koopmann et al. 2001). It seems that the CO emission is also slightly stronger in the northern arm than the southern arm.
}
\label{}
\end{figure}

\begin{figure}
\begin{center}
\end{center}
\caption{Same as figure 2 for NGC 4689. Contour levels of the CO integrated intensity map are 3 $\times$ (1, 2, 3, 4) K km s$^{-1}$. Contour levels of the velocity field are from 1530 to 1690 km s$^{-1}$ with an interval of 20 km s$^{-1}$. Contour levels of the P -- V diagram are 0.08 and 0.16 K. The peak of CO is offset from the center of the galaxy. Even with an interferometer, the central concentration of molecular gas is not apparent (Sofue et al. 2003a).
}
\label{}
\end{figure}

\clearpage

\begin{figure}
\begin{center}


\end{center}
\caption{Same as figure 2 for NGC 4736. Contour levels of the CO integrated intensity map are 4 $\times$ (1, 2, 3, 4, 6, 8) K km s$^{-1}$. Contour levels of the velocity field are from 220 to 420 km s$^{-1}$ with an interval of 20 km s$^{-1}$. Contour levels of the P -- V diagram are 0.1 $\times$ (1, 2, 3, 4) K. A tightly wound spiral structure is seen in the CO map of BIMA SONG (Regan et al. 2001) and the mid-IR map (Roussel et al. 2001). Our map also shows a similar structure with a strong peak at the center. 
}
\label{}

\end{figure}

\begin{figure}
\begin{center}
\end{center}
\caption{Same as figure 2 for NGC 5055. Contour levels of the CO integrated intensity map are 6 $\times$ (1, 2, 3, 4, 6, 8, 10) K km s$^{-1}$. Contour levels of the velocity field are from 330 to 660 km s$^{-1}$ with an interval of 30 km s$^{-1}$. Contour levels of the P -- V diagram are 0.15 $\times$ (1, 2, 3, 4) K. The data were taken from Kuno et al. (1997). Weak two-arm spiral structure is seen in the CO map, which coincides with the spiral arms seen in $K$ band (Thornley 1996; Kuno et al. 1997; Thornley, Mundy 1997). NGC 5055 has the most prominent central CO peak of all SA galaxies in our sample. The rotation curve rises steeply at the center. 
}
\label{}
\end{figure}

\begin{figure}
\begin{center}
\end{center}
\caption{Same as figure 2 for NGC 5194. Contour levels of the CO integrated intensity map are 5 $\times$ (1, 2, 3, 4, 6, 8, 10, 15, 20) K km s$^{-1}$. Contour levels of the velocity field are from 400 to 540 km s$^{-1}$ with an interval of 20 km s$^{-1}$. Contour levels of the P -- V diagram are 0.15 $\times$ (1, 2, 3, 4, 6, 8, 10) K. The data were taken from Nakai et al. (1994). Grand design spiral arms are seen in the CO map. See Nakai et al. (1994), Kuno et al. (1995), and Kuno and Nakai (1997) for details.
}

\label{}
\end{figure}

\begin{figure}
\begin{center}
\end{center}
\caption{Same as figure 2 for NGC 5236. Contour levels of the CO integrated intensity map are 8 $\times$ (1, 2, 3, 4, 6, 8, 10, 15, 20) K km s$^{-1}$. Contour levels of the velocity field are from 450 to 570 km s$^{-1}$ with an interval of 20 km s$^{-1}$. Contour levels of the P -- V diagram are 0.15 $\times$ (1, 2, 3, 4, 6, 8, 10) K. The P-V diagram along the major axis shows the rigid-like rotation curve in the bar region. This is due to the same reason as for UGC 2855. That is, the radial velocity corresponds to the pattern speed of the bar, since the line of sight is nearly perpendicular to the bar (Kuno et al. 2000).
}

\label{}
\end{figure}

\begin{figure}
\begin{center}
\end{center}
\caption{Same as figure 2 for NGC 5248. Contour levels of the CO integrated intensity map are 3 $\times$ (1, 2, 3, 4, 6, 8, 10, 15) K km s$^{-1}$. Contour levels of the velocity field are from 1070 to 1270 km s$^{-1}$ with an interval of 20 km s$^{-1}$. Contour levels of the P -- V diagram are 0.1 $\times$ (1, 2, 3, 4) K. A large-scale bar structure was found by Jogee et al. (2002). The CO emission is detected at the center and along the spiral arms in the bar (or the offset ridges along the large-scale bar). CO emission is not detected in the spiral arms. 

}
\label{}
\end{figure}

\begin{figure}
\begin{center}
\end{center}
\caption{Same as figure 2 for NGC 5247. Contour levels of the CO integrated intensity map are 2 $\times$ (1, 2, 3, 4, 5, 6, 7) K km s$^{-1}$. Contour levels of the velocity field are from 1320 to 1400 km s$^{-1}$ with an interval of 20 km s$^{-1}$. Contour levels of the P -- V diagram are 0.07 $\times$ (1, 2, 3, 4) K. The open spiral arms can be traced in the CO map. The southern arm is longer than the northern arm in optical images. The southern molecular arm of CO is also longer and more prominent than the northern arm.
}
\label{}
\end{figure}

\begin{figure}
\begin{center}
\end{center}

\caption{Same as figure 2 for NGC 5457. Contour levels of the CO integrated intensity map are 3 $\times$ (1, 2, 3, 4, 6) K km s$^{-1}$. Contour levels of the velocity field are from 180 to 320 km s$^{-1}$ with an interval of 10 km s$^{-1}$. Contour levels of the P -- V diagram are 0.1 $\times$ (1, 2, 3, 4) K. The bar-like structure in the central region shown by Kenney, Scoville, \& Wilson (1991) can be seen. The peak at the center is not so prominent. Although optical images of NGC 5457 show multiple arms, two spiral arms are prominent in CO, starting from the ends of the bar-like structure. The molecular arms coincide with arms seen in the near-IR band. There are some CO peaks which correspond to the HII regions. The giant HII region NGC 5461 is most prominent in our CO map. It is interesting that while tracers of star forming regions, such as H$\alpha$ (Cedr\'es, Cepa 2002), far-UV (Bianchi et al. 2005), and mid-IR (Roussel et al. 2001) are stronger in the eastern side than in the western side, the distribution of the molecular gas is opposite.
}
\label{}
\end{figure}

\begin{figure}
\begin{center}

\end{center}
\caption{Same as figure 2 for NGC 6217. Contour levels of the CO integrated intensity map are 5 $\times$ (1, 2, 3, 4, 6) K km s$^{-1}$. Contour levels of the velocity field are from 1320 to 1380 km s$^{-1}$ with an interval of 20 km s$^{-1}$. Contour levels of the P -- V diagram are 0.1 $\times$ (1, 2, 3) K. CO emission is detected at the center and along the bar. 
}
\label{}
\end{figure}

\begin{figure}
\begin{center}
\end{center}
\caption{Same as figure 2 for NGC 6946. Contour levels of the CO integrated intensity map are 5 $\times$ (1, 2, 3, 4, 6, 8, 10, 15, 20, 30, 40) K km s$^{-1}$. Contour levels of the velocity field are from -20 to 140 km s$^{-1}$ with an interval of 20 km s$^{-1}$. Contour levels of the P -- V diagram are 0.15 $\times$ (1, 2, 3, 4, 6, 8) K. Our map shows quite similar distribution of molecular gas with that of CO(2--1) obtained with the IRAM 30-m telescope (Walsh et al. 2002). Compared with the very strong peak at the center, the disk emission is fairy weak. The spiral arms are not prominent in CO, although there are some peaks in the arms. The peaks correspond to bright HII regions.
}
\label{}
\end{figure}

\begin{figure}
\begin{center}
\end{center}
\caption{Same as figure 2 for NGC 6951. Contour levels of the CO integrated intensity map are 3 $\times$ (1, 2, 3, 4, 6, 8, 10) K km s$^{-1}$. Contour levels of the velocity field are from 1310 to 1580 km s$^{-1}$ with an interval of 30 km s$^{-1}$. Contour levels of the P -- V diagram are 0.07 $\times$ (1, 2, 3) K. There is a strong CO peak at the center. Strong concentrations of molecular gas are seen at the bar ends. While CO emission is detected in spiral arms, the emission is weak in the bar region, where H$\alpha$ is also weak (M\'arquez, Moles 1993). An elongated structure from the central region toward the northern arm is seen.
}
\label{}
\end{figure}

\clearpage

\begin{figure}
\begin{center}
\FigureFile(80mm,50mm){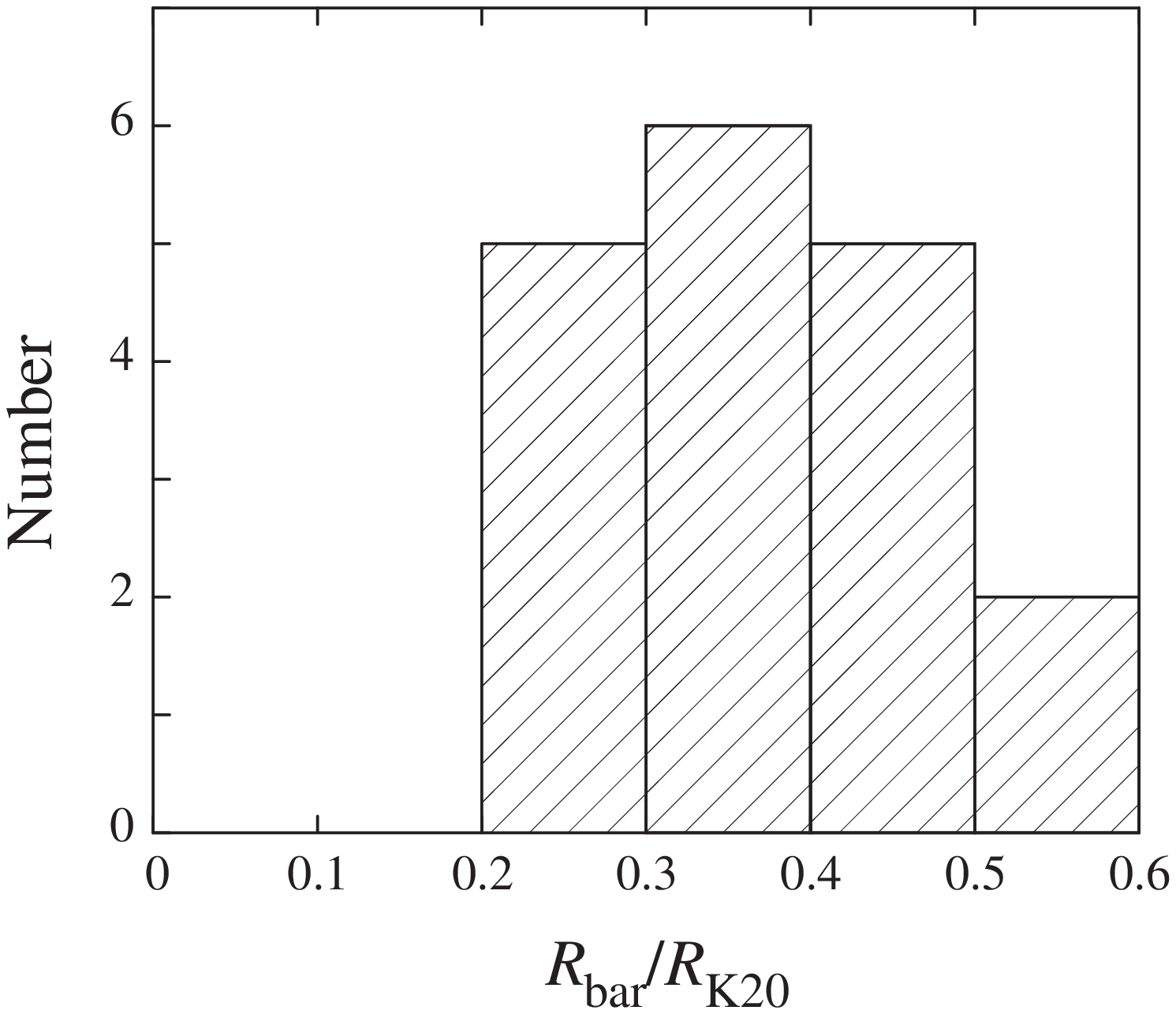}
\end{center}
\caption{Histgram of $R_{\rm bar}/R_{\rm K20}$.}
\label{}
\end{figure}

\begin{figure}
\begin{center}
\end{center}
\caption{Histgram of $f_{\rm in}$ for SA and SAB+SB galaxies.}
\label{}
\end{figure}

\begin{figure}
\begin{center}
\FigureFile(80mm,50mm){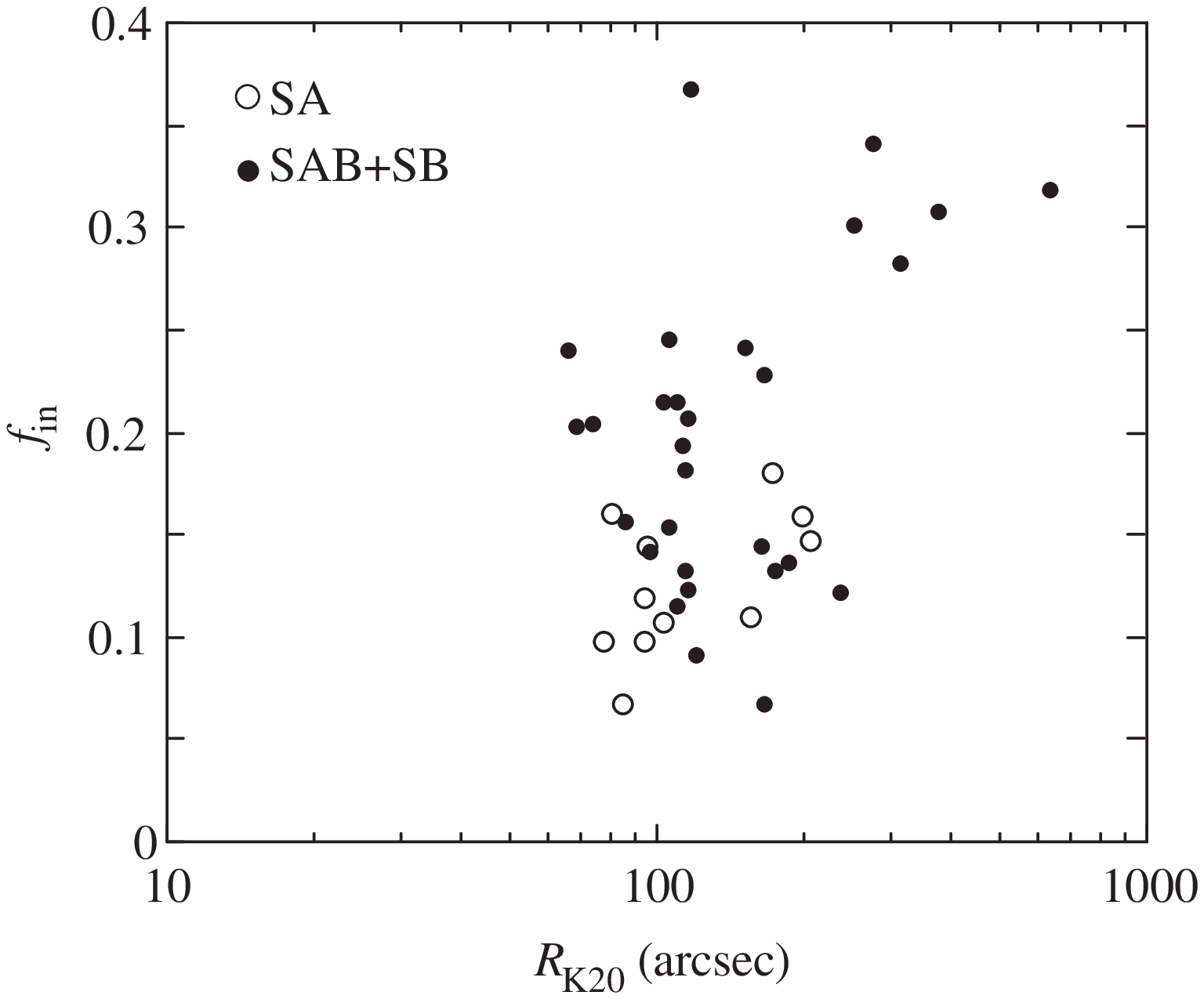}

\end{center}
\caption{Comparison of $f_{\rm in}$ with $R_{\rm K20}$. Open and solid circles indicate SA and SAB+SB galaxies, respectively. }

\label{}
\end{figure}

\begin{figure}
\begin{center}

\FigureFile(80mm,50mm){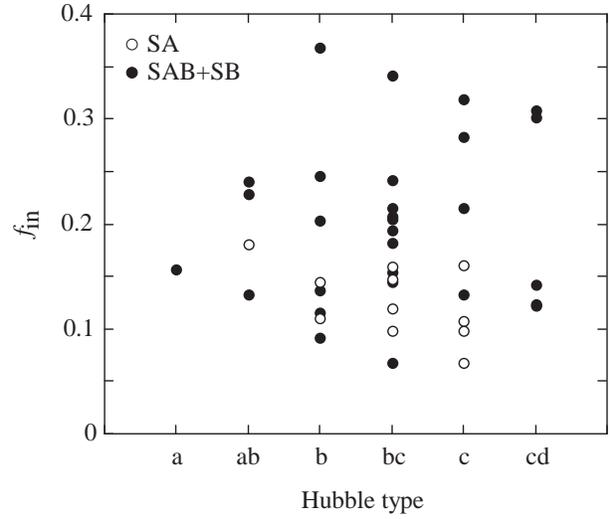}
\end{center}
\caption{Comparison of $f_{\rm in}$ with Hubble type. Open and solid circles indicate SA and SAB+SB galaxies, respectively. }
\label{}
\end{figure}

\begin{figure}
\begin{center}
\FigureFile(80mm,50mm){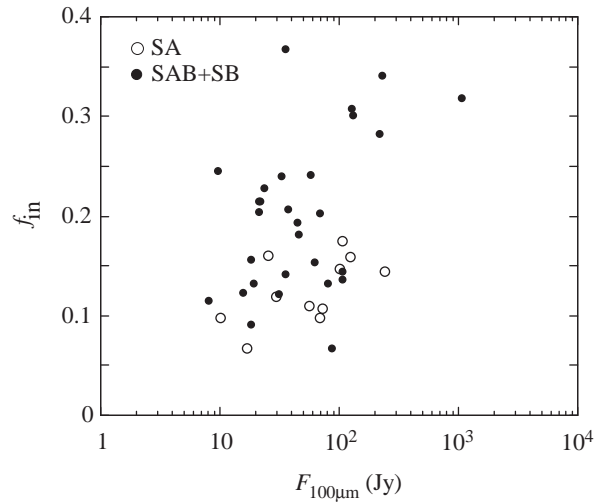}
\end{center}
\caption{Comparison of $f_{\rm in}$ with flux at 100 $\mu$m. Open and solid circles indicate SA and SAB+SB galaxies, respectively. }
\label{}

\end{figure}

\begin{figure}
\begin{center}
\end{center}
\caption{Histgram of $f_{\rm out}$ for SA and SAB+SB galaxies.}
\label{}
\end{figure}

\begin{figure}
\begin{center}
\FigureFile(80mm,50mm){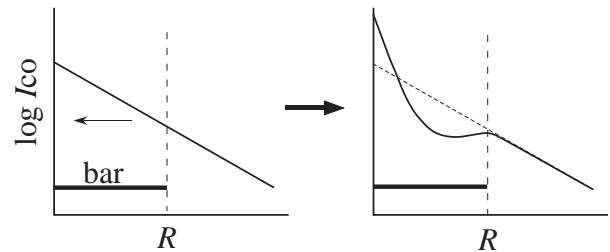}
\end{center}
\caption{Schematic diagram of the radial distribution of molecular gas in barred spiral galaxies.}
\label{}

\end{figure}

\begin{figure}
\begin{center}
\FigureFile(80mm,50mm){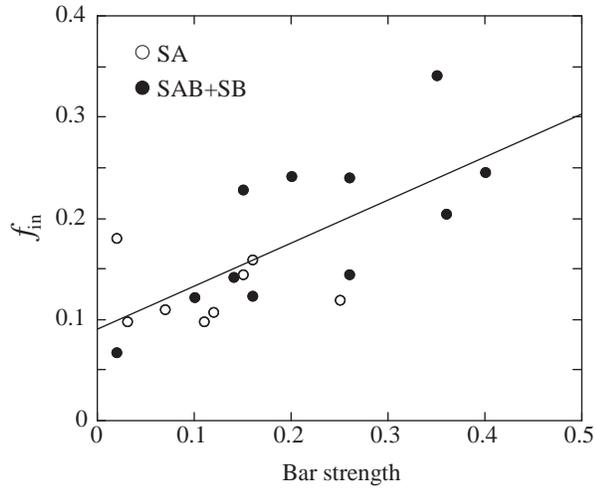}
\end{center}
\caption{Comparison of $f_{\rm in}$ with bar strength which is $Q_{H}$ in Laurikainen and Salo (2002). Open and solid circles indicate SA and SAB+SB galaxies, respectively. The solid line is the best fit to the data. }
\label{}
\end{figure}

\begin{figure}
\begin{center}
\FigureFile(80mm,50mm){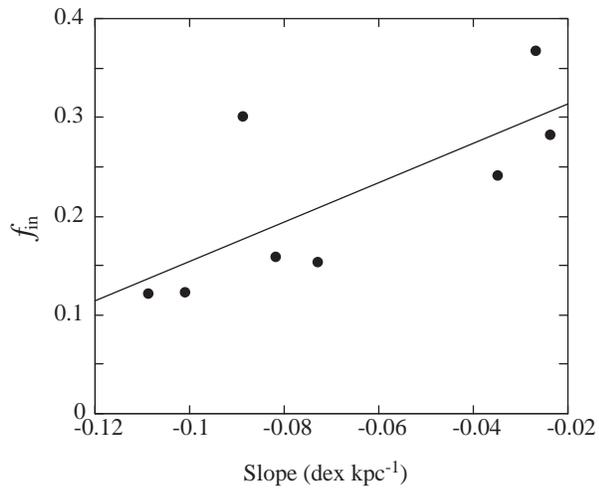}
\end{center}
\caption{Comparison of $f_{\rm in}$ with the abundance gradient which is referred from Martin and Roy (1994). The solid line is the best fit to the data. }
\label{}
\end{figure}


\begin{longtable}{ccccccccc}
\caption{Observational parameters}
\hline\hline
Name & R.A. & Dec. & Ref.\footnotemark[$*$] & Grid spacing\footnotemark[$\dagger$] & Grid number\footnotemark[$\ddagger$] & P.A. \footnotemark[$\S$] & Receiver\footnotemark[$\|$] & Ref.\footnotemark[$\#$] \\
  & (J2000) & (J2000) &  & (arcsec) &  & (deg) &  &  \\
\hline
\endhead

\hline
\endfoot
\hline
\multicolumn{9}{l}{\hbox to 0pt{\parbox{170mm}{\footnotesize 
\par\noindent
\footnotemark[$*$] References of the coordinate of the center. 1: Keto et al. (1993) 2: Ishiguro et al. (1989) 3: Capetti et al. (1997) 4: Falco et al. (2000) 5: 2MASS EXTENDED OBJECTS. Final Release (2003) 6: History and Accurate Positions for the NGC/IC Objects (Corwin, H. G. 2004). The catalogue can be queried from http://vizier.u-strasbg.fr/viz-bin/VizieR?-source=VII/239A. 7: Condon et al. (1990) 8: Dressel, Condon (1976) 9: Eck et al. (2002) 10: Jarrett et al. (2003) 11: Sofue et al. (2003b) 12: Takase, Miyauchi (1989) 13: Becker, White \& Helfand (1995) 14: van der Kruit (1973) 15: Ford et al. (1985) 16: Sukumar et al. (1987) 17: the NASA/IPAC Extragalactic Database (NED) 18: Veron-Cetty, Veron (1996) 19: Green (1994) 20: Kenney et al. (1992)
\par\noindent
\footnotemark[$\dagger$] Grid spacing of the mapping.
\par\noindent
\footnotemark[$\ddagger$] Number of the observed points.
\par\noindent
\footnotemark[$\S$] Position angle of the coordinate of the grids.
\par\noindent
\footnotemark[$\|$]  1: S100 2: S115Q 3: BEARS.
\par\noindent
\footnotemark[$\#$] Reference of CO data. 1: Sorai et al. (2000a) 2: This work 3: Kuno et al. (2000) 4: Nakanishi (2005) 5: Kuno et al. (1997) 6: Nakai et al. (1994) 
}\hss}}
\endlastfoot
NGC 253 & $\timeform{00h47m33s.22}$ & $-\timeform{25D17'16".19}$ & 1 & 17 & 576 & 51 & 2 & 1 \\
Maffei 2 & $\timeform{02h41m54s.6}$ & $+\timeform{59D36'12".3}$ & 2 & 11 & 576 & 26 & 2 & 2 \\
NGC 1068 & $\timeform{02h42m40s.711}$ & $+\timeform{00D00'47".81}$ & 3 & 10.3 & 576  & 89 & 3 & 2 \\
IC342 & $\timeform{03h46m48s.9}$ & $+\timeform{68D05'46".0}$ & 4 & 10.3 & 4480 & 37   & 3 & 2 \\
UGC 2855 & $\timeform{03h48m20s.73}$ & $+\timeform{70D07'58".4}$ & 5 & 10.3 & 1056 & 109 & 3 & 2 \\
NGC 2903 & $\timeform{09h32m10s.11}$ & $+\timeform{21D30'03".0}$ & 5 & 10.3 & 1056 & 12.5 & 3 & 2 \\
NGC 3184 & $\timeform{10h18m16s.98}$ & $+\timeform{41D25'27".8}$ & 5 & 10.3 & 576  & 135 & 3 & 2 \\
NGC 3351 & $\timeform{10h43m57s.73}$ & $+\timeform{11D42'13".0}$ & 6 & 10.3 & 576  & 13 & 3 & 2 \\
NGC 3504 & $\timeform{11h03m11s.24}$ & $+\timeform{27D58'21".2}$ & 7 & 11 & 72 & 149  & 2 & 3 \\
NGC 3521 & $\timeform{11h05m48s.88}$ & $-\timeform{00D02'15".04}$ & 8 & 10.3 & 384  & 166 & 3 & 2 \\
NGC 3627 & $\timeform{11h20m15s.027}$ & $+\timeform{12D59'29".58}$ & 9 & 10.3 & 1056 & 176 & 3 & 2 \\
NGC 3631 & $\timeform{11h21m02s.65}$ & $+\timeform{53D10'16".5}$ & 7 & 10.3 & 576  & 150 & 3 & 2 \\
NGC 4051 & $\timeform{12h03m09s.7}$ & $+\timeform{44D31'52".5}$ & 10 & 10.3 & 576  & 131 & 3 & 2 \\
NGC 4102 & $\timeform{12h06m23s.12}$ & $+\timeform{52D42'39".4}$ & 5 & 10.3 & 600  & 38 & 3 & 2 \\
NGC 4192 & $\timeform{12h13m48s.29}$ & $+\timeform{14D54'01".9}$ & 11 & 10.3 & 1056 & 155 & 3 & 2 \\
NGC 4212 & $\timeform{12h26m39s.4}$ & $+\timeform{13D54'04".6}$ & 11 & 10.3 & 576  & 75 & 3 & 4 \\
NGC 4254 & $\timeform{12h18m49s.61}$ & $+\timeform{14D24'59".6}$ & 11 & 10.3 & 1056 & 68 & 3 & 2 \\
NGC 4303 & $\timeform{12h21m54s.67}$ & $+\timeform{04D28'19".83}$ & 7 & 11 & 144 & 138 & 2 & 2 \\
NGC 4321 & $\timeform{12h22m54s.92}$ & $+\timeform{15D49'19".84}$ & 12 & 11 & 303  & 153 & 2 & 2 \\
NGC 4402 & $\timeform{12h26m07s.4}$ & $+\timeform{13D06'44".7}$ & 11 & 10.3 & 576  & 90 & 3 & 4 \\
NGC 4414 & $\timeform{12h26m27s.1}$ & $+\timeform{31D13'24".7}$ & 5 & 10.3 & 576 & 158 & 3 & 2 \\
NGC 4419 & $\timeform{12h26m56s.4}$ & $+\timeform{15D02'50".2}$ & 11 & 10.3 & 576  & 133 & 3 & 4 \\
NGC 4501 & $\timeform{12h31m59s.17}$ & $+\timeform{14D25'13".7}$ & 5 & 11 & 413 & 142 & 2 & 2 \\
NGC 4535 & $\timeform{12h34m20s.35}$ & $+\timeform{08D11'52".2}$ & 11 & 10.3 & 576  & 0 & 3 & 2 \\
NGC 4536 & $\timeform{12h34m27s.08}$ & $+\timeform{02D11'17".1}$ & 11 & 10.3 & 576  & 125 & 3 & 2 \\
NGC 4548 & $\timeform{12h35m26s.44}$ & $+\timeform{14D29'47".4}$ & 11 & 10.3 & 576  & 136 & 3 & 4 \\
NGC 4569 & $\timeform{12h36m49s.82}$ & $+\timeform{13D09'45".8}$ & 11 & 10.3 & 1056 & 23 & 3 & 2 \\
NGC 4579 & $\timeform{12h37m43s.53}$ & $+\timeform{11D49'05".5}$ & 11 & 10.3 & 576  & 95 & 3 & 2 \\
NGC 4654 & $\timeform{12h43m56s.67}$ & $+\timeform{13D07'36".1}$ & 11 & 10.3 & 576  & 125 & 3 & 2 \\
NGC 4689 & $\timeform{12h50m15s.86}$ & $+\timeform{13D29'27".4}$ & 11 & 10.3 & 576  & 166 & 3 & 4 \\
NGC 4736 & $\timeform{12h50m53s.061}$ & $+\timeform{41D07'13".65}$ & 13 & 10.3 & 576  & 115 & 3 & 2 \\
NGC 5055 & $\timeform{13h15m49s.34}$ & $+\timeform{42D01'45".57}$ & 14 & 11 & 505  & 103 & 2 & 5 \\
NGC 5194 & $\timeform{13h29m52s.7}$ & $+\timeform{47D11'42".60}$ & 15 & 7.5, 15 &  768 & -10 & 1 & 6 \\
NGC 5236 & $\timeform{13h37m00s.48}$ & $-\timeform{29D51'56".48}$ & 16 & 11 & 730  & 45 & 2 & 2 \\
NGC 5248 & $\timeform{13h37m32s.07}$ & $+\timeform{08D53'06".2}$ & 5 & 10.3 & 576  & 105 & 3 & 2 \\
NGC 5247 & $\timeform{13h38m03s.03}$ & $-\timeform{17D53'03".4}$ & 5 & 10.3 & 576  & 108 & 3 & 2 \\
NGC 5457 & $\timeform{14h03m12s.48}$ & $+\timeform{54D20'55".3}$ & 17 & 10.3 & 2296 & 44 & 3 & 2 \\
NGC 6217 & $\timeform{16h32m38s.59}$ & $+\timeform{78D11'52".2}$ & 18 & 11 & 54 & 122 & 2 & 2 \\
NGC 6946 & $\timeform{20h34m52s.336}$ & $+\timeform{60D09'14".21}$ & 19 & 10.3 & 1443 & 60 & 3 & 2 \\
NGC 6951 & $\timeform{20h37m14s.27}$ & $+\timeform{66D06'21".7}$ & 20 & 11 & 238 & 157 & 2 & 2 \\
\end{longtable}

\begin{table}
\begin{center}
\caption{Distribution of galaxy types of this sample}
\begin{tabular}{ccccccc}
\hline\hline

Hubble type & a & ab & b & bc & c & cd \\
\hline
Number & 1 & 4 & 8 & 14 & 8 & 5 \\
\hline\hline
\multicolumn{2}{c}{Arm Class} & \multicolumn{2}{c}{1 -- 4} & \multicolumn{2}{c}{5 -- 12} \\
\hline
\multicolumn{2}{c}{Number} & \multicolumn{2}{c}{8} & \multicolumn{2}{c}{26} \\
\hline\hline
bar/non-bar & \multicolumn{2}{c}{SA} & \multicolumn{2}{c}{SAB} & \multicolumn{2}{c}{SB} \\
\hline
Number & \multicolumn{2}{c}{11} & \multicolumn{2}{c}{24} & \multicolumn{2}{c}{4} \\

\hline
\end{tabular}
\end{center}
\end{table}

\end{document}